\documentstyle[12pt]{article}
\newcommand{\be}{\begin{equation}}
\newcommand{\ee}{\end{equation}}
\begin{document}

\begin{center}
{\bf Chiral Solutions to Generalized Burgers and Burgers-Huxley
Equations\footnote{This work is supported in part by funds
provided by the U. S. Department of Energy (D.O.E.) under
cooperative research agreement DE-FC02-94ER40818, and by Conselho
Nacional de Desenvolvimento Cient\'\i fico e Tecnol\'ogico, CNPq, Brazil.}}
\end{center}

\begin{center}
D. Bazeia\footnote{On leave from Departamento de F\'\i sica, Universidade
Federal da Para\'\i ba, Caixa Postal 5008, 58051-970 Jo\~ao Pessoa,
Para\'\i ba, Brazil}
\end{center}

\begin{center}
Center for Theoretical Physics\\
Laboratory for Nuclear Science and Department of Physics\\
Massachusetts Institute of Technology, Cambridge, Massachusetts 02139-4307
\end{center}

\vskip 1.5cm

\begin{center}
(MIT-CTP-2714, February 1998)
\end{center}

\vskip 1cm

\begin{center}
Abstract
\end{center}

We investigate generalizations of the Burgers and Burgers-Huxley equations.
The investigations we offer focus attention mainly on presenting explict
analytical solutions by means of relating these generalized equations to
relativistic $1+1$ dimensional systems of scalar fields where topological
solutions are known to play a role. Emphasis is given on chiral solutions,
that is, on the possibility of finding solutions that travel with velocities
determined in terms of the parameters that identify the generalized equation,
with a definite sign.

\vskip 1.5cm

\begin{center}
PACS numbers: 03.40.Kf; 47.27.Eq; 83.10.Ji
\end{center}

\newpage

\section{Introduction}
Recent works \cite{ben96,jac96} have shown that the nonlinear Schr\"odinger
equation presents chiral solitons when nonlinearity enters the game via
derivative coupling. The nonlinear Schr\"odinger equation that appears in
the above works is obtained from a very interesting dimensional reduction to
one space dimension of a planar model \cite{jpi90} describing non-relativistic
matter coupled to a Chern-Simons gauge field \cite{jac95}. This equation can
be cast to the form
\begin{equation}
iu_t+\lambda\,j\,u+\mu\,u_{xx}=\frac{dV}{d\rho}\,u~,
\end{equation}
where $u_t=\partial u/\partial t$, $u_{xx}=\partial^2u/\partial x^2$,
$j=-i\nu(u^{*}u_x -u\,u^{*}_x)$ is the current density, $\lambda,\mu,\nu$
are real parameters and $V=V(u^*u)=V(\rho)$ is the potential, expressed in
terms of the number density. This equation should be contrasted with
\begin{equation}
iu_t+\mu\,u_{xx}=\frac{dV}{d\rho}\,u~,
\end{equation}
which is the standard nonlinear Schr\"odinger equation -- recall that
one usually considers $V(\rho)$ as quadratic or cubic in $\rho$.

For travelling waves in the form $u(x,t)=\rho(x-ct)\exp[i\theta(x,t)]$ one
finds solutions to the above nonlinear derivative Schr\"odinger equation that
present velocity restricted to just one sense, and the system is then chiral.
This is nicely illustrated in a more recent work on the same subject
\cite{gse97}, where the soliton structure for vanishing and non-vanishing
boundary conditions are investigated. Another very recent work \cite{his97}
also finds chiral solitons in a derivative multi-component nonlinear
Schr\"odinger equation. Evidently, the chiral solitons found in these works
may play important role within the context of the fractional quantum Hall
effect, where chiral excitations are known to appear.

Inspired by these former works, another investigation \cite {bmo98}
was recently done, and there it was shown how to extend the presence of chiral
solitons to the context of generalized Korteweg-de Vries or gKdV equations.
The gKdV equation considered in that work can be cast to the form
\be
u_t+A\,g(u)\,u_x-\delta\,u_{xxx}=0~,
\ee
where $g(u)=g(-u)$ is a smooth function and $A$ and $\delta$ are real
parameters. Here there are travelling waves in the form $u(x,t)=u(x-ct)$
that also engender chiral behavior, with the velocity $c$ determined in terms
of the parameters that define the generalized equation, with a definite sign.
The method offered in \cite{bmo98} is based on the possibility of mapping the
problem of searching for soliton solutions to gKdV equations to some known
problem, related to the presence of topological solutions in relativistic $1+1$
dimensional systems of real scalar fields. This issue is somehow direct in
this case since for travelling waves the gKdV equation becomes an ordinary
differential equation that presents two terms, one containing the third
derivative of the configuration with respect to $y=x-ct$, and the other
having first derivative. In this case a first integral leads to a equation
that relates the second order derivative of the configuration to a given
function of this very same configuration. This function depends on the
particular generalized equation one is considering, but the equation has
exactly the form of the equation of motion for static field configuration
describing relativistic $1+1$ dimensional systems of real scalar field.

In the present paper we shall extend the above idea to other scenarios,
qualitatively different from the one that appears related to the generalized
KdV equation. Here we shall consider generalized Burgers and Burgers-Huxley
equations. As one knows, the Burgers equation \cite{bur48} was first
introduced to describe turbulence in one space dimension, and has been used
in several other physical contexts \cite{bur74}, including for instance
sound waves in viscous media. The Burgers-Huxley equation \cite{deb97} is
more complicated and makes the system qualitatively different from the former
Burgers equation, and we shall need distinct approaches to deal with each one.
We shall be mainly concerned with providing explicit solutions to different
models, governed by distinct functions that specify systems where
the generalized Burgers and Burgers-Huxley equations play a role. Within
this context, it is interesting to realize that all the systems that we
investigate have the general feature of being generalizations of Burgers
and Burgers-Huxley equations, for which we only relax the assumption
of weak nonlineary. This means that we are not changing the original equations
to introduce other effects, like including a new term to describe dispersion
for instance, but just changing the nonlinear properties of the original
system, described by the Burgers or the Burgers-Huxley equation. For the
generalized Burgers equations, for instance, we consider the dynamics of
diffusion in media where nonlinearity is not just restricted to the
simplest case. In particular, we introduce the modified Burgers equation,
e. g. the equation that is obtained from the standard Burgers equation
by just changing its nonlinear term in exactly the same way one changes the
nonlinear term in the KdV equation to get to the modified KdV or mKdV equatiom.

The investigations focus attention mainly on finding solutions that engender
the same chiral behavior recently found in nonlinear derivative Schr\"odinger
equations \cite{ben96,jac96,gse97}, and in generalized KdV equations
\cite{bmo98}. As we are going to show in the next sections, we succeed in
presenting several distinct and nontrivial systems and some explicit solutions
engendering chiral behavior. The chiral solutions we have found are
travelling solitary waves and we name them chiral waves. Since we want to
map the above problems to problems related to relativistic $1+1$
dimensional systems of real scalar fields, in the next
Sec.~{\ref{sec:fields}} we introduce the main results for the relativistic
systems that we shall need in the rest of the paper. In Sec.~{\ref{sec:bur}}
we investigate generalized Burgers equations and coupled generalized Burgers
equations, which are second order differential equations. We deal with
further generalizations in Sec.~{\ref{sec:bhu}}, where we investigate
generalized Burgers-Huxley equations and coupled generalized Burgers-Huxley
equations.  In this case the generalized equations follows from the
generalized Burgers equations by adding another term, non-derivative,
and this makes the former approach, valid for generalized Burgers equations,
not to work anymore. In spite of this, however, we have been able to map
generalized Burgers-Huxley equations to first order equations that are
related to systems of scalar fields. Furthermore, we could yet extend these
equations to the case of two coupled equations, describing two configurations
interacting with each other in a very specific way. We end the paper in
Sec.~{\ref{sec:com}}, where we introduce comments and conclusions.

\section{Relativistic Systems of Real Scalar fields}
\label{sec:fields}

Our main motivation in this section is to comment on relativistic $1+1$
dimensional systems of real scalar fields. We do this by focusing attention on
the respective equations of motion and on some interesting properties that
appear when specific classes of systems are considered. Further details can be
found in Ref.~{\cite{raj82}}, for instance. For simplicity we split
the subject in the several subsections that follow, all of them related to
issues that appear in the rest of this work.

\subsection{Systems with one field}
\label{sec:1field}

Let us start by considering the Lagrangian density
\be
{\cal L}=\frac{1}{2}\frac{\partial\phi}{\partial x_{\alpha}}\,
\frac{\partial\phi}{\partial x^{\alpha}}-V(\phi)~,
\ee
where $x^{\alpha}=(x^0=t,x^1=x)$, $x_{\alpha}=(x_0=t,x_1=-x)$, and
$V(\phi)$ is the potential that defines the specific system we want to
consider. There is a class of systems that is identified by potentials
defined by
\be
V(\phi)=\frac{1}{2}\left(\frac{dH}{d\phi}\right)^2~,
\ee
where $H=H(\phi)$ is a smooth but otherwise arbitrary function of the
field $\phi$. This specific form of the potential is very interesting because
it allows introducing the following result: the equation of motion that follows
from the above Lagrangian density can be written in the form, for static field
$\phi=\phi(x)$,
\be
\frac{d^2\phi}{dx^2}=\frac{dH}{d\phi}\frac{d^2H}{d\phi^2}~,
\ee
and it is also solved by field configurations that solve the
following first order differential equation
\be
\frac{d\phi}{dx}=\frac{dH}{d\phi}~.
\ee
This result is important because we can deal with the above first order
differential equation to write solutions to the equation of motion,
which is second order.

The standard examples of systems of one real scalar field are the $\phi^4$,
the $\phi^6$ and the sine-Gordon systems, which can be written in terms of the
following functions, respectively,
\be
H(\phi)=\lambda a^2 \phi -\frac{1}{3}\lambda \phi^3~,
\ee

\be
H(\phi)=\frac{1}{2}\lambda a^2\phi^2-\frac{1}{4}\lambda\phi^4~,
\ee
and
\be
H(\phi)=-\lambda\,\cos(\phi).
\ee

\subsection{Systems with two fields}
\label{sec:2field}

For systems of two fields we consider the Lagrangian density
\be
{\cal L}=\frac{1}{2}\frac{\partial\phi}{\partial x_{\alpha}}\,
\frac{\partial\phi}{\partial x^{\alpha}}+
\frac{1}{2}\frac{\partial\chi}{\partial x_{\alpha}}\,
\frac{\partial\chi}{\partial x^{\alpha}}-V(\phi,\chi)~.
\ee
Here $V(\phi,\chi)$ is the potential that defines the specific system of two
fields we want to consider. There is a class of systems that is identified
by potentials defined via \cite{baz95}
\be
V(\phi,\chi)=\frac{1}{2}\left(\frac{\partial H}{\partial\phi}\right)^2+
\frac{1}{2}\left(\frac{\partial H}{\partial\chi}\right)^2~,
\ee
where $H=H(\phi,\chi)$ is a smooth but otherwise arbitrary function of the
two fields $\phi$ and $\chi$. This specific form of the potential is very
interesting because it allows introducing the result that the equations of
motion that follow from the above Lagrangian density are, for static fields
$\phi=\phi(x)$ and $\chi=\chi(x)$,
\begin{eqnarray}
\frac{d^2\phi}{dx^2}&=&\frac{\partial H}{\partial\phi}
\frac{\partial^2H}{\partial\phi^2}+\frac{\partial H}{\partial\chi}
\frac{\partial^2H}{\partial\chi\partial\phi}~,\\
\frac{d^2\chi}{dx^2}&=&\frac{\partial H}{\partial\phi}
\frac{\partial^2H}{\partial\phi\partial\chi}+\frac{\partial H}{\partial\chi}
\frac{\partial^2H}{\partial\chi^2}~.
\end{eqnarray}
and they are also solved by field configurations that obey the
following set of first order differential equations, as it was recently
realized in \cite{baz95},
\begin{eqnarray}
\frac{d\phi}{dx}&=&\frac{\partial H}{\partial\phi}~,\\
\frac{d\chi}{dx}&=&\frac{\partial H}{\partial\chi}~.
\end{eqnarray}
This result is important because we can deal with the above pair of first
order differential equation to write solutions to the equations of motion,
which are second order.

To quote explicit examples of systems of two fields we introduce
\be
H(\phi,\chi)=\lambda a^2\phi-\frac{1}{3}\lambda\phi^3-\mu\phi\chi^2~,
\ee
and 
\be
H(\phi,\chi)=\frac{1}{2}\lambda a^2\phi^2-\frac{1}{4}\lambda\phi^4-
\frac{1}{2}\mu\phi^2\chi^2~,
\ee
which define systems recently investigated in \cite{baz95}.

\subsection{Other properties}

Since we are working in $1+1$ dimensions we can introduce the following vector
\be
j^{\alpha}=\epsilon^{\alpha\beta}\partial_{\beta}H~,
\ee
where $H$ is $H(\phi)$ or $H(\phi,\chi)$, depending on which of the two
above classes of systems we are dealing with. Now, thanks to symmetry
properties of the Levi-Civita tensor, the above vector obeys
$\partial_{\alpha}j^{\alpha}=0$, which is a continuity equation, and so
$j^{\alpha}$ can be seen as a conserved current density. This conserved
current is topological because it is directly related to topological
properties of the vacuun manifold of the corresponding Lagrangian system.

To see how topological properties enter the game we recall that the above
current allows introducing the conserverd charge
$Q=H(x=\infty)-H(x=-\infty)$, which makes sense if and only if
$H(x=\infty)$ is different from $H(x=-\infty)$; however, one knows that $H$
depends on $x$ via field configurations, and so $\phi$, in the case of one
field, or $(\phi,\chi)$, in the case of two fields, must become distinct
asymptotic points in the line $\phi$ or in the plane $(\phi,\chi)$,
respectively. On the other hand, in order to make the energy finite one can
only accept field configurations that become vacuum states asymptoticaly,
and so we can only have nontrivial solutions when the vacuum manifold
contains at least two distinct points. When this is the case, the conserved
charge $Q=H(x=\infty)-H(x=-\infty)$ is known as the topological charge
associated to the corresponding topological solution.

Since the potential is defined in terms of squares of derivative of $H$ we
see that the vacuum states are given by $dH/d\phi=0$, in the case of systems
of one field, and by $\partial H/\partial\phi=0$ and
$\partial H/\partial\chi=0$ in the case of systems of two fields. These
identifications allow introducing an interesting connection to dynamical
systems, that appears very naturally in the above classes of systems of real
scalar fields because these systems satisfy first order differential equations,
and these first order equations are naturally seen as dynamical systems.

If we see the first order equations as dynamical systems, it is
straightforward to recognize that the singular points are nothing but all the
vacuum states in the corresponding classes of relativistic systems of real
scalar fields, and this helps searching for solutions of the relativistic
systems. To see how this works explicitly, let us recall that we can classify
the singular points standardly, and this strongly helps in the identification
of topological sectors and of orbits that connect the singular points or vacuum
states asymptoticaly. The picture that then emerges is very important for
searching for explicit solutions, since now we can use the trial orbit method
first introduced in \cite{raj79} in a much more effective way -- see the works
\cite{baz95} for several explicit examples.

\section{Generalized Burgers Equations}
\label{sec:bur}

The Burgers equation \cite{bur48}
\be
\label{eq:bur}
u_t+A\, u\, u_x-\nu\, u_{xx}=0~,
\ee
can be seen as a reduction of the Navier-Stokes equation to the case of a
single space dimension. In this equation $A$ controls the nonlinearity
and $\nu$ stands for viscosity. This equation was first introduced to describe
turbulence in a single space dimension \cite{bur74}. It is perhaps the
simplest nonlinear differential second order equation, and it has been
considered to describe different physical problems such as sound waves in
viscous media and magnetohydrodynamic waves in media with finite electrical
conductivity.

In the present work, however, instead of considering the Burgers equation
we shall deal with generalized Burgers equations that describe a single
configuration $u(x,t)$ and two interacting configurations $u(x,t)$ and
$v(x,t)$. For simplicity we split the subject into the two subsections that
follow.

\subsection{Systems with one equation}

Insteady of considering the Burgers Eq.~{$(\ref{eq:bur})$}, let us deal
with the equation
\be
\label{eq:gbur}
u_t+g(u)\,u_x-\nu\, u_{xx}=0~,
\ee
where $g(u)$ is a smooth function of $u$. This is the generalized form of
the Burgers equation, and the Burgers equation $(\ref{eq:bur})$
is obtained with the linear function $g(u)=A\,u$. Like the Burgers equation,
the generalized Burgers equation $(\ref{eq:gbur})$ also combines nonlinearity
and diffusion, but now nonlinearity is controlled by $g(u)$ and may vary
according to the model one considers -- note that the Burgers equation
is defined with the simplest nontrivial function $g=g(u)$.

The present investigation is inspired in a former work \cite{bmo98}, and here
we follow the route already introduced therein, where generalizations of the
KdV equation have been considered. To do this we focus attention
to generalized Burgers equations defined with even functions $g(u)=g(-u)$.
In this case the generalized equation $(\ref{eq:gbur})$ presents the discrete
$Z_2$ symmetry $u\to-u$, which will be further explored below. Since our main
motivation is to find explicit travelling waves, let us search for
configuration in the form $u(x,t)=u(x-ct)= u(y)$. In this case
Eq.~{(\ref{eq:gbur})} reduces to
\be
\label{eq:gbur1}
\frac{du}{dy}=-\frac{c}{\nu}\,u+\frac{1}{\nu}\,\int^{u}\,du'\,g(u')~,
\ee
if we require that the resulting equation does not break the discrete symmetry
$u\to-u$. We now introduce another function $h=h(u)$ even in $u$
in order to rewrite the above equation as
\be
\label{eq:gbur11}
\frac{du}{dy}=\frac{dh}{du}~,
\ee
and this implies that
\be
-\frac{c}{\nu}\, u+\frac{1}{\nu}\int^u\,du'\,g(u')=\frac{dh}{du}~.
\ee
Such a result is interesting because we can use it to relate solutions
to the generalized Burgers equation to field configurations that describe
relativistic $1+1$ dimensional systems, as we have already introduced in
Sec.~{\ref{sec:1field}}.

It is interesting to remark that the generalized Burgers equation
$(\ref{eq:gbur})$ is a second order differential equation, and so the
resulting equation $(\ref{eq:gbur1})$ is necessarily first order, and this
shows that when the potential of the corresponding field system
is written in the specific form $V(\phi)=(1/2)({dH}/{d\phi})^2$ we can
relate the generalized equation $(\ref{eq:gbur})$ to relativistic $1+1$
dimensional system defined by this potential. The situation here is different
from the case considered in \cite{bmo98}, since there the generalized KdV
equation is directly mapped to second order equations of motion of field
systems.

Before presenting explicit examples, let us first recognize that we are only
considering $h(u)$ even in $u$. Then, for polynomial functions the fourth or
higher power on $u$ must be necessarily present, and this fact
eliminates the possibility of mapping the generalized Burgers equation
$(\ref{eq:gbur})$ to the popular $\phi^4$ model, since in this case one would
require an odd function. However, in the case where
\be
H(\phi)=\frac{1}{2}\lambda a^2 \phi^2-\frac{1}{4} \lambda \phi^4~,
\ee
we get to the $\phi^6$ model, that presents the first order equation
\be
\frac{d\phi}{dx}=\lambda a^2\phi-\lambda\phi^3~,
\ee
which has the kink solutions
\be
\phi_{\pm}(x)=\pm\sqrt{\frac{1}{2}a^2\{1\pm\tanh[\lambda a^2(x-\bar{x})]\}\,}~,
\ee
where $\bar{x}$ is arbitrary.

These solutions are also solutions to the corresponding generalized Burgers
equation. To see this explicitly, let us consider the generalized Burgers
equation
\be
u_t+3A\,u^2\, u_x-\nu\, u_{xx}=0~.
\ee
Here $A$ and $\nu$ are real parameters. We remark that we are using
$g(u)=3\,A\,u^2$, and this compared to the Burgers equation (for which
$g(u)=Au$) is perhaps the next simplest case of nonlinearity. This equation
may be named the modified Burgers equation, since it contains nothing but the
change $u\to u^2$ in its nonlinear term, and this is exactly what happens
with the mKdV equation, which incorporates the same modification in relation
to the standard KdV equation.

In the present case the first order equation corresponding to travelling wave
$u=u(y)$ can be cast to the form
\be
\frac{du}{dy}=-\frac{c}{\nu}\, u+\frac{A}{\nu}\, u^3~.
\ee
Here we see that if $A$ and $c$ have the same sign we can get several
distinct cases. As an example let us supposse that $c>0$, $B>0$ and $\nu>0$:
in this case we can write the solution, for instance,
\be
u(x,t)=\sqrt{\frac{1}{2}\left(\frac{c}{A}\right)\Biggl[1-
\tanh\left(\frac{c}{\nu}\right)(x-ct-\bar{x})\Biggr]\,\,}~.
\ee
We see that the velocity is restricted to obey $c>0$, and this makes the
solution chiral.

Another example is related to the sine-Gordon system, and in this case the
travelling solution present a single velocity. Here we consider the generalized
equation
\be
\label{eq:bsg}
u_t+[A+\lambda\,\cos(B\,u)]\,u_x-\nu\,u_{xx}=0~.
\ee
$A$, $B$, $\nu$ and $\lambda$ are all real parameters. Like in the former
cases, we search for solutions that obey $u(x,t)=u(x-ct)=u(y)$ to get to
\be
\frac{du}{dy}=\frac{\lambda}{\nu\,B}\,\sin(B\,u)~,
\ee
after setting $c-A=0$. This is just the equation that appears in the
sine-Gordon system, and has the soliton solution
\be
u(x,t)=\frac{2}{B}\arctan e^{\frac{\lambda}{\nu}\,(x-A\,t-\bar{x})}~,
\ee
which represents chiral solutions, with velocity given by the parameter $A$
that appears in the original equation $(\ref{eq:bsg})$.

As we have seen, the generalized Burges equation is a second order
differential equation, and our approach leads to first order equations that
can be used to map the equations of motion of relativistic $1+1$ dimensional
systems {\it after} some fine-tuning on the velocity of the travelling
solution is introduced. This fine-tuning in general depends on the parameters
that defines the generalized Burgers equation, and there are situations where
it restricts the velocity to just a single value and sense.

\subsection{Systems with two coupled equations}

The generalized Burgers Eq.~{$(\ref{eq:gbur})$} can be rewritten in the form
\be
u_t+f_x-\nu\, u_{xx}=0~,
\ee
and for $f=f(u)$ we get
\be
u_t+\frac{df}{du}\,u_x-\nu\, u_{xx}=0~.
\ee
This form is interesting since it allows a natural extension to systems where
two or more configurations interact with each other. See Ref.~{\cite{his97}}
for similar considerations in the case of the derivative nonlinear
Schr\"odinger equation. In the case of two interacting configurations
$u=u(x,t)$ and $v=v(x,t)$ we have
\begin{eqnarray}
u_t+f_x-\nu\, u_{xx}&=&0~,\\
v_t+g_x-\bar{\nu}\, v_{xx}&=&0~.
\end{eqnarray}
For $f=f(u,v)$ and $g=g(u,v)$ we can write
\begin{eqnarray}
u_t+\frac{\partial f}{\partial u}\,u_x+
\frac{\partial f}{\partial v}\,v_x-\nu\, u_{xx}&=&0~,\\
v_t+\frac{\partial g}{\partial u}\,u_x+
\frac{\partial g}{\partial v}\,v_x-\bar{\nu}\, v_{xx}&=&0~.
\end{eqnarray}
Here we note that the above system of equations presents the symmetry
$u\to-u$ and $v\to-v$ in $(u,v)$ space when $f(u,v)$ is odd in $u$ and
even in $v$, and $g(u,v)$ in even in $u$ and odd in $v$. This is the case we
consider in this subsection.

Like in the former case, let us search for travelling waves
$u(x,t)=u(x-ct)=u(y)$ and $v(x,t)=v(x-ct)=v(y)$. Here we get to the equations
\begin{eqnarray}
\frac{du}{dy}&=&-\frac{c}{\nu}\,u+\frac{1}{\nu}\,f(u,v)~,\\
\frac{dv}{dy}&=&-\frac{c}{\bar{\nu}}\,v+\frac{1}{\bar{\nu}}\,g(u,v)~,
\end{eqnarray}
if one requires that the same symmetry $u\to-u$ and $v\to-v$ present in the
original system of equations appears in the above equations. We now introduce
another function $h=h(u,v)$ to rewrite the above first order equations
in the form
\begin{eqnarray}
\frac{du}{dy}&=&\frac{\partial h}{\partial u}~,\\
\frac{dv}{dy}&=&\frac{\partial h}{\partial v}~,
\end{eqnarray}
and this implies
\begin{eqnarray}
-\frac{c}{\nu}\,u+\frac{1}{\nu}\,f(u,v)&=&
\frac{\partial h}{\partial u}~,\\
-\frac{c}{\bar{\nu}}\,v+\frac{1}{\bar{\nu}}\,g(u,v)&=&
\frac{\partial h}{\partial v}~.
\end{eqnarray}
This result naturally leads to relativistic systems of coupled scalar fields,
as already introduced in Sec.~{\ref{sec:2field}}. We note that $h(u,v)$
must be even in both $u$ and $v$, in order to preserve the symmetry properties
of $f(u,v)$ and $g(u,v)$.

As an illustration, let us consider the pair of equations
\begin{eqnarray}
u_t+\nu(3\lambda u^2+\mu v^2)\,u_x+
2\mu\nu uvv_x-\nu u_{xx}&=&0~,\\
v_t+(A+\mu\bar{\nu}u^2)v_x+2\mu\bar{\nu}uvu_x-\bar{\nu}v_{xx}&=&0~,
\end{eqnarray}
where $A$, $\lambda$, $\mu$, $\nu$ and $\bar{\nu}$ are real parameters. For
travelling waves $u=u(y)$ and $v=v(y)$ we get
\begin{eqnarray}
\frac{du}{dy}&=&-\frac{c}{\nu}\,u+\lambda u^3+\mu u v^2~,\\
\frac{dv}{dy}&=&-\frac{c-A}{\bar{\nu}}\,v+\mu u^2 v~.
\end{eqnarray}
This system maps systems of coupled scalar fields recently investigated, that
can be used to give solutions to the above pair of generalized equations. For
$c-A=0$ see for instance \cite{bmo98,baz95} and the next section.  For
$c-A\ne0$ we use results of Ref.~{\cite{etb97}} to give another explicit
example, which presents the pair of solutions, for instance,
\begin{eqnarray}
u(x,t)&=&\frac{1}{2}\sqrt{1-\tanh[({c}/{\nu})(x-ct-\bar{x})]\,\,}~,\\
v(x,t)&=&\frac{1}{2}\sqrt{(1/2)\,[1-
\tanh[({c}/{\nu})(x-ct-\bar{x})]\,\,}~,
\end{eqnarray}
provided that $\lambda=c/\nu$, $\mu=2(c-A)/\bar{\nu}$ and
$c=A\nu/(\nu-\bar{\nu})$. Here it is interesting to see that for
$\bar{\nu}\to\nu$ this system presents chiral solutions of the form
introduced above only in the limit $A\to0$.

\section{Generalized Burgers-Huxley Equations}
\label{sec:bhu}

The Burgers-Huxley equation \cite{deb97}
\be
u_t+A\,u\,u_x -\nu\,u_{xx}=B\,u\,(1-u)(u-\gamma)~,
\ee
modifies the Burgers equation by just adding the non-derivative term given
above. It is used as a nonlinear model to investigate wave propagation mainly
in biological and chemical systems. This equation was already investigated for
instance in \cite{sat87,wan90}, and in \cite{oth75} one finds examples
related to biological systems where travelling waves play a role.
In the present section, however, we consider two distinct and new
generalizations of the above equation, and for this reason we split the
subject in two subsection. In the next subsection we deal with generalized
Burgers-Huxley equations that describe systems where a single configuration
$u=u(x,t)$ is present. In the other subsection we investigate coupled
generalized Burgers-Huxley equations that describe systems where two
configurations $u(x,t)$ and $v(x,t)$ interact with each other.

\subsection{Systems with one equation}

In this section we further extend the generalized Burgers equation to the
following form
\be
\label{eq:bhu}
u_t+\frac{dg}{du}\,u_x-\nu\,u_{xx}=B\,f(u)~.
\ee
which represents the generalized form of the Burgers-Huxley equation that we
consider in the present work. Here $B$ and $\nu$ are real parameters, and
$f=f(u)$ and $g=g(u)$ are smooth functions. We note that for $B=0$ we get
back to the generalized Burgers equation investigated in Sec.~{\ref{sec:bur}},
and for $g(u)$ the trivial constant function we get to the generalized
Fisher equation \cite{bri86}. We see that the added non-derivative contribution
makes the former approach not to work since a first integral can not be
trivially done anymore.

Like in the former Sec.~{\ref{sec:bur}}, however, we search for travelling
solutions in the form $u=u(y)$ to get
\be
\left(-c+\frac{dg}{du}\right)\frac{du}{dy}-
\nu\frac{d^2u}{dy^2}=B\,f(u)~.
\ee
The important result here is that this equation is solved by
\be
\frac{du}{dy}=\frac{1}{\nu}\,g(u)~,
\ee
if one sets $c=-\nu\,B$ and $f(u)=g(u)$. This is very interesting since we can
relate the above Burgers-Huxley equation to relativistic $1+1$ dimensional
systems of scalar fields, and so we can get different equations and solutions
given in terms of different functions g=g(u). Here we remark that every
solution we get in this approach is chiral since it is constrained to
obey the condition $c=-\nu\,B$.

To give an explicit example let us get inspiration on the $\phi^6$ model,
that is, let us consider the following generalized Burgers-Huxley equation
\be
u_t+(A-3\,u^2)u_x-\nu\,u_{xx}=B\, u\,(A-u^2)~.
\ee
We name this equation the modified Burgers-Huxley equation, and here we
search for travelling solutions in the form $u(y)$ to get to
\be
(-c+A-3\,u^2)\frac{du}{dy}-\nu\,\frac{d^2u}{dy^2}=
B\, u\,(A-u^2)~,
\ee
which is solved by the first order equation
\be
\frac{du}{dy}=\frac{1}{\nu}\,u\,(A-u^2)~.
\ee
if one sets $c=-\nu\,B$. For $A>0$ an explicit solution is, for instance,
\be
u(x,t)=\sqrt{(A/2)\,[1+\tanh(A/\nu)\,(x+\nu\,B\,t-\bar{x})]\,\,}~,
\ee
and present chiral behavior.

\subsection{Systems with two coupled equations}

The generalized Burgers-Huxley equation $(\ref{eq:bhu})$ can be further
generalized to the case where several configurations interact with each
other. For simplicity we consider the case where two confugurations $u(x,t)$
and $v(x,t)$ are present. The point here is that the generalized Burgers-Huxley
equation admits a natural extension to the following pair of coupled
Burgers-Huxley equations
\begin{eqnarray}
u_t+\frac{\partial g}{\partial u}\,u_x+\frac{\partial g}{\partial v}\,v_x-
\nu\,u_{xx}&=&B\,g(u,v)~,\\
v_t+\frac{\partial\bar{g}}{\partial v}\,v_x+
\frac{\partial\bar{g}}{\partial u}\,u_x-\bar{\nu}\,v_{xx}
&=&\bar{B}\,\bar{g}(u,v)~.
\end{eqnarray}
For solutions in the form of travelling waves $u=u(y)$ and $v=v(y)$  we get
\begin{eqnarray}
\left(-c+\frac{\partial g}{\partial u}\right)\frac{du}{dy}+
\frac{\partial g}{\partial v}\,\frac{dv}{dy}-
\nu\,\frac{d^2u}{dy^2}&=&B\,g(u,v)~,\\
\left(-c+\frac{\partial\bar{g}}{\partial v}\right)\frac{dv}{dy}+
\frac{\partial\bar{g}}{\partial u}\,\frac{du}{dy}-
\bar{\nu}\,\frac{d^2v}{dy^2}&=&\bar{B}\,\bar{g}(u,v)~.
\end{eqnarray}
These equations are solved by configurations obeying the first order equations
\begin{eqnarray}
\frac{du}{dy}&=&\frac{1}{\nu}\,g(u,v)~,\\
\frac{dv}{dy}&=&\frac{1}{\bar{\nu}}\,\bar{g}(u,v)~,
\end{eqnarray}
if one imposes $c=-\nu\,B=-\bar{\nu}\,\bar{B}$, and this is the way we get to
relativistic systems of coupled real scalar fields. Here we can introduce
another function $h(u,v)$ in order to rewrite the above equations in the form
\begin{eqnarray}
\frac{du}{dy}&=&\frac{\partial h}{\partial u}~,\\
\frac{dv}{dy}&=&\frac{\partial h}{\partial v}~,
\end{eqnarray}
which implies
\begin{eqnarray}
\frac{1}{\nu}\,g(u,v)&=&\frac{\partial h}{\partial u}~,\\
\frac{1}{\bar{\nu}}\,\bar{g}(u,v)&=&\frac{\partial h}{\partial v}~,
\end{eqnarray}
and this reproduces the first order equations that defines the class of
systems of two coupled fields considered in Sec.~{\ref{sec:2field}}.

To give an explicit example, let us consider $h(u,v)$ given by
\be
h(u,v)=\frac{1}{2}\lambda A\,u^2-\frac{1}{4}\lambda\,u^4-\frac{1}{2}
\mu\,u^2\,v^2~.
\ee
Here $A$, $\lambda$ and $\mu$ are real parameters. In this case the
generalized Burgers-Huxley equations are
\begin{eqnarray}
u_t+\nu\,(\lambda A-3\lambda u^2-\mu v^2)u_x-2\,\mu\,\nu\,u\,v\,v_x-
\nu\,u_{xx}&=&B\,\nu\,u\,(\lambda A-\lambda\,u^2-\mu\,v^2)~,\\
v_t-\mu\,\bar{\nu}\,u^2\,v_x-2\mu\,\bar{\nu}\,u\,v\,u_x-\bar{\nu}\,v_{xx}
&=&-\bar{B}\,\bar{\nu}\,\mu\,u^2\,v~.
\end{eqnarray}
For travelling waves $u=u(y)$ and $v=v(y)$ we get
\begin{eqnarray}
\left(-\frac{c}{\nu}+\lambda A-3\lambda u^2-
\mu v^2\right)\,\frac{du}{dy}-2\,\mu\,u\,v\,\frac{dv}{dy}-\frac{d^2u}{dy^2}
&=&B\,u\,(\lambda A-\lambda\,u^2-\mu\,v^2)~,\\
\left(\frac{c}{\bar{\nu}}+\mu\,u^2\right)\,\frac{dv}{dy}+
2\mu\,u\,v\,\frac{du}{dy}+\frac{d^2v}{dy^2}&=&\bar{B}\,\mu\,u^2\,v~.
\end{eqnarray}
For $c=-\nu\,B=\bar{\nu}\bar{B}$ the above equations are solved by the
following pair of first order differential equations
\begin{eqnarray}
\frac{du}{dy}&=&\lambda\,u\,(A-u^2)-\mu\,u\,v^2~,\\
\frac{dv}{dy}&=&-\mu\,u^2\,v~.
\end{eqnarray}
This system maps the system already investigated in \cite{baz95}, and 
an interesting pair of solutions is, for instance, for $A>0$,
\begin{eqnarray}
u(x,t)&=&\sqrt{(A/2)[1+\tanh\mu A(x-ct-\bar{x})]\,\,}~,\\
v(x,t)&=&\sqrt{(A/2)(\lambda/\mu-1)[1-\tanh\mu A(x-ct-\bar{x})]\,\,}~,
\end{eqnarray}
which is valid for $\lambda/\mu>1$ and for $c=-\nu\,B=-\bar{\nu}\bar{B}$.
Here we see that the above pair of first order equations also presents
the solutions
\be
u(x,t)=\sqrt{(A/2)[1+\tanh\lambda A(x+\nu\,B\,t-\bar{x})]\,\,}~,
\ee
and $v(x,t)=0$. It is interesting to see that this pair of solutions
can be obtained from the former pair of solutions by considering
$\mu\to\lambda$, and that it reproduces the solution found in the former
system, described by a single generalized Burgers-Huxley equation.
 
\section{Comments and Conclusions}
\label{sec:com}

In the present paper we have investigated generalized Burgers and
Burgers-Huxley equations. These investigations are inspired in a former work
\cite{bmo98}, in which generalizations of the Korteweg-de Vries equation have
been considered. Although the Burgers and KdV equations are different, the
difference relies essentially on the presence of the second order derivative
term that responds for diffusion in the Burgers equation and the third order
derivative term that controls dispersion in the KdV equation.

In Ref.~{\cite{bmo98}} it was shown that a first integral of the KdV equation
leads to a second order equation that may directly map the equation of
motion for relativistic $1+1$ dimensional system of real scalar fields. The
same approach applies to the Burgers equation, but now one is led to a first
order differential equation, and so we can only map the equation of motion for
relativistic scalar systems indirectly, via solutions to first order
differential equations. In the sequel, we investigated generalized
Burgers-Huxley equations. In this case the generalized equations follow from
the generalized Burgers equations by introducing non-derivative contributions,
and these contributions make the former approach, valid for generalized Burgers
equations, not to work anymore. In spite of this, however, we have been able
to map generalized Burgers-Huxley equations to first order equations that
are related to systems of scalar fields. Furthermore, we could yet
extend these equations to the case of two coupled equations, describing
two configurations interacting with each other in a very specific way.
Several examples have been introduced in Sec.~{\ref{sec:bur}} and in
Sec.~{\ref{sec:bhu}}, to illustrate the general procedure and to present
explicit solutions exibiting chiral behavior.

As a general feature of the investigations offered in the present work,
we have shown that solutions to the generalized equations can be found as
solutions of first order differential equations, in the case of one
configuration $du/dy=dh/du$, and in the case of two configurations
$du/dy={\partial h}/{\partial u}$ and $dv/dy={\partial h}/{\partial v}$.
We have used this fact to link solutions to these generalized equations to
topological solutions of the related system of one or two real scalar fields.
However, there are other routes, one of them being related to the fact that
these first order equations can be directly mapped to continuous
population models for one specie and for two interacting species, respectively.
As one knows, population models have been extensively studied in theoretical
biology \cite{mur93}, and so we are unvieling another way of finding new
solutions to generalized Burgers and Burgers-Huxley equations.

There are other issues that deserve further investigations, for instance the
one that relies on searching for other nonlinear equations and finding
corresponding chiral solutions. Since we have already investigated
generalized KdV and Burgers equations, perhaps the most natural case that now
appears is related to generalized equations that add together the Burgers
and KdV equations. As one knows, investigations concerning dynamics of
weak nonlinearity in the presence of both diffusion and dispersion was
inniciated in Ref.~{\cite{joh70}}, and we shall show in a future work
\cite{baz98} that there exists an interesting generalized KdV-Burgers
equation that in the limit of vanishing dispersion gives a generalized
Kardar-Parisi-Zhang equation \cite{kpz86} that may perhaps help introducing
new models for evolution of the profile of growing interfaces.

\vskip 1cm

We would like to thank Roman Jackiw for interesting comments, and for reading
the manuscript. We also thank Henrique Boschi-Filho and Fernando Moraes for
discussions and for interest in the present work.

\end{document}